\documentclass
[aps,prl,amsfonts,amssymb,twocolumn,amsmath,preprintnumbers,nofootinbib,floatfix,superscriptaddress]{revtex4-1}%
\usepackage[dvips]{graphics}
\usepackage{graphicx}
\usepackage{bm}
\usepackage{amsmath}
\usepackage{amsfonts}
\usepackage{amssymb}
\usepackage{xcolor}
\usepackage{subfigure}
\usepackage{hyperref,hypcap}
\usepackage{braket}
\usepackage{commath}%
\providecommand{\U}[1]{\protect\rule{.1in}{.1in}}
\begin{document}

\title{Conserved current of nonconserved quantities}
\author{Cong Xiao}
\affiliation{Department of Physics, The University of Texas at Austin, Austin, Texas 78712, USA}
\affiliation{Department of Physics, The University of Hong Kong, Hong Kong, China}
\affiliation{HKU-UCAS Joint Institute of Theoretical and Computational Physics at Hong Kong, China}
\author{Qian Niu}
\affiliation{Department of Physics, The University of Texas at Austin, Austin, Texas 78712, USA}

\begin{abstract}
We provide a unified semiclassical theory for the conserved current of
nonconserved quantities, and manifest it in two physical contexts: the spin
current of Bloch electrons and the charge current of mean-field Bogoliubov
quasiparticles. Several longstanding problems that limit the playground of
the conserved spin current of electrons are solved. We reveal that the hitherto
overlooked torque quadrupole density and Berry phase correction to the
torque dipole density are essential to assure a circulating spin current with
vanishing net flow at equilibrium. The band geometric origin of bulk spin
transport is ascertained to be the momentum space spin texture and Berry curvature instead of the spin Berry
curvature, paving the way for material related studies.
In superconductors the attained conserved charge current corresponds to
the quasiparticle charge current renormalized by the condensate backflow.
Its circulation at equilibrium gives an orbital magnetization, which involves
the characteristics of superconductivity, such as the Berry
curvature arising from unconventional pairing and an orbital magnetic moment
induced by the charge dipole of moving quasiparticles.
\end{abstract}
\maketitle

%\author{Cong Xiao}
%\affiliation{Department of Physics, The University of Texas at Austin, Austin, Texas 78712, USA}
%\author{Qian Niu}
%\affiliation{Department of Physics, The University of Texas at Austin, Austin, Texas 78712, USA}

\emph{{\color{blue} Introduction.}}---In condensed matter physics the current
of a nonconserved quantity is intriguing, as the conventionally
defined current as the anticommutator of that quantity and velocity is in
general non-circulating with a nonvanishing net current flow through a cross
section of a sample even at equilibrium, such as the spin current of
spin-orbit coupled Bloch electrons \cite{Rashba2003} and the charge current of
mean-field superconducting quasiparticles \cite{Nambu2009}. There has not been
a unified recipe showing a conserved current whose circulation characterizes
the corresponding orbital magnetization and whose net flow vanishes at equilibrium.

In spintronics research, conserved spin currents have been investigated
\cite{Murakami2004,Sun2005,Shi2006,Murakami2006,Guo2009,Sugimoto2006,Mandal2008,Nagaosa2018,Xiao2018,Freimuth2010,Gorini2012,Mele2019,Culcer2004}%
. A natural way towards a bulk conserved current is to include the current
density due to the source term of the continuity equation, and a conserved
spin current was attempted along this line \cite{Shi2006,Murakami2006,Guo2009}%
. However, that proposal cannot address equilibrium currents of the pivotal
role, hence the spin orbital magnetization characterizing the circulating
conserved spin current has not been touched on and whether or not
the net flow vanishes at equilibrium is unclear. Without a knowledge about
the spin orbital magnetization, it is unknown what the transport component of
the conserved current is in the presence of statistical forces, i.e.,
gradients of chemical potential and temperature, as the circulating current
should be discounted to obtain the transport one
\cite{Cooper1997,Streda1977,Xiao2006}. Even in the case of
electrically induced transport, the band geometric origin of the conserved
current remains to be unveiled, and a conductivity formula amenable to a
momentum space electronic structure code is absent \cite{Turek2019}. These
unknowns severely limit the utility of the conserved current in spintronics
studies \cite{Sinova2015}.

In the context of superconductivity, although charge is ultimately conserved,
that of bare mean-field Bogoliubov quasiparticles is not, as these
quasiparticles are not eigenstates of charge \cite{Nambu1960}. The
conventional charge current of such quasiparticles is not conserved
\cite{Nambu2009}, with the source term arising from the charge transfer
between quasiparticles and condensate
\cite{Nambu1960,BTK1982,Kivelson2012,Wu2017}. How to understand the current
due to this source term in a semiclassical description of quasiparticles is
quite elusive, and the orbital magnetization as a circulating conserved charge
current was not addressed yet. Moreover, the connection and
possible unification of this subject and the conserved spin current of Bloch
electrons, in which the nonconservation related symmetry breaking are
respectively spontaneous and explicit, have not been studied.

In this Letter we present a unified semiclassical theory at steady states for
the conserved current of nonconserved internal degrees of freedom (denoted by
operator $\hat{\boldsymbol{s}}$). We uncover that the attained current has a
vanishing net flow at equilibrium and acquire the orbital magnetization of
nonconserved quantities. In the context of superconductivity, we recognize
that this conserved current corresponds to a semiclassical description of the
charge current renormalized by the condensate backflow due to the coupling
between quasiparticles and condensate \cite{Nambu1960,Kivelson2012}. A Berry
phase formula for the orbital magnetization in superconductors is found, which
consists of both the local and global charge circuits of quasiparticles
with distinct geometric origins. The global circuit is due to the
momentum space Berry curvature derivable from not only the parent Bloch states
but also the unconventional superconducting pairing, whereas the local one
includes a nontrivial orbital magnetic moment induced by the charge dipole
moment of a moving quasiparticle, which has never been seen for charge
conserved particles.

The aforementioned problems on the conserved spin current of Bloch electrons
are solved. In particular, the torque quadrupole density and a
Berry phase contribution to the torque dipole density are found to be vital to
ensure a circulating spin current with vanishing net flow at equilibrium. The
band structure dictated spin transport is expressed by the momentum space spin
texture and Berry curvature, instead of the so called spin Berry curvature for
the conventional spin current \cite{Flatte2015,Yao2005,Yan2016}. To get to
this result, we extend the notion of coordinate shift (side jump)
\cite{Sinitsyn2006,Sinitsyn2008} to the shift of a spin coordinate.
The established formulas for time-reversal even and odd conserved
spin currents pave the way for material related studies.

\emph{{\color{blue} Equilibrium current.}}---For the convenience of
presentation we describe the theory in terms of Bloch electrons. In view of
the continuity equation $\boldsymbol{\nabla}\cdot\boldsymbol{J}%
^{\boldsymbol{s}}\left(  \boldsymbol{r}\right)  =\boldsymbol{\tau}\left(
\boldsymbol{r}\right)  $ at steady states, in order to construct a conserved
current of $\hat{\boldsymbol{s}}$ we need study the local density of the
conventional $\boldsymbol{s}$-current, represented by the operator
$\boldsymbol{\hat{J}}^{\boldsymbol{s}}=\frac{1}{2}\left\{  \boldsymbol{\hat
{v}},\boldsymbol{\hat{s}}\right\}  $ with $\boldsymbol{\hat{v}}$ being the
velocity, and that of the $\boldsymbol{s}$-generation rate, represented by
$\boldsymbol{\hat{\tau}}=d\boldsymbol{\hat{s}}/dt$. As we concern current
densities up to the first order of spatial gradients in order to capture the
magnetization current, pursuing $\boldsymbol{\tau}(\boldsymbol{r})$ up to the
second order is indispensable. To this end we develop a second order
semiclassical theory in combination with the field variational approach
\cite{Dong2020}. The derivation of this theory is involved hence delegated to
the Supplemental Material \cite{supp}, whereas the results take physically
transparent forms and are set forth below.

The conventional $\boldsymbol{s}$-current density at equilibrium up to the
first order of spatial gradients reads
\begin{equation}
J^{i\boldsymbol{s}}=\int f_{n}J_{n}^{i\boldsymbol{s}}-\partial_{l}%
D^{li\boldsymbol{s}},\text{ }D^{li\boldsymbol{s}}=\int(f_{n}d_{n}%
^{li\boldsymbol{s}}+g_{n}\Omega_{n}^{li\boldsymbol{s}}).
\label{conventional current}%
\end{equation}
The first term in the zeroth order of gradients is in general nonzero
\cite{Rashba2003}, and the second term in the first order may not be
circulating, in contrast to the orbital magnetization current which is always
circulating \cite{Cooper1997}. Therefore, the net flow of this
conventional $\boldsymbol{s}$-current through a cross section of a sample may
not vanish even at equilibrium. Thus, this current alone cannot afford a
physical description of the flow of nonconserved quantities. The
non-circulating nature of the gradient term in Eq. (\ref{conventional current}%
) originates from the nonconservation of $\boldsymbol{s}$ and follows directly
from the fact that the tensor $D^{li\boldsymbol{s}}$, which is the equilibrium
dipole density of the conventional $\boldsymbol{s}$-current \cite{Dong2020},
is not antisymmetric with respect to Cartesian indices $i$ and $l$ for such
$\boldsymbol{s}$. Here $d_{n}^{lis}=${{$\operatorname{Re}\langle\left(
\hat{r}-r_{c}\right)  ^{l}\hat{J}^{i\boldsymbol{s}}\rangle$}} is the dipole
moment of the distribution of the conventional $\boldsymbol{s}$-current on the
spread of a wave packet constructed by superposing Bloch states from a
particular band $n$, and $\boldsymbol{r}_{c}$ denotes the probability center
of the wave packet. $\Omega_{n}^{li\boldsymbol{s}}=-2\operatorname{Im}%
\sum_{n_{1}\neq n}v_{nn_{1}}^{l}J_{n_{1}n}^{i\boldsymbol{s}}/\omega_{nn_{1}%
}^{2}$ is the so called spin Berry curvature when $\boldsymbol{s}$ is spin
\cite{Yao2005,Yan2016}, and is thus termed as the $\boldsymbol{s}$-Berry
curvature in the following. A similar expression for
$J^{i\boldsymbol{s}}$ appeared in Ref. \cite{Culcer2004}, but missed the
$\boldsymbol{s}$-Berry curvature contribution to the total dipole density. If
$\boldsymbol{s}$ is conserved, $\int f_{n}J_{n}^{i\boldsymbol{s}}$ vanishes,
whereas $D^{li\boldsymbol{s}}$ becomes antisymmetric and proportional to the
orbital magnetization \cite{Xiao2010}, hence $J^{i\boldsymbol{s}}$ reduces to
a circulating magnetization current with vanishing net flow.

The notation $\int$ without integral variable is shorthand for $\int\left[
d\boldsymbol{k}\right]  $, and $\left[  d\boldsymbol{k}\right]  \equiv\sum
_{n}d\boldsymbol{k}/\left(  2\pi\right)  ^{D}$, where $\boldsymbol{k}$ is the
crystal momentum ($\hbar=1$), $D$ is the spatial dimensionality. $f_{n}$ is
the Fermi function, and $g_{n}=k_{B}T\ln(1-f_{n})$ is the state resolved grand
potential density, with $T$ as the temperature. $v_{nn_{1}}^{l}=\langle
u_{n}|\hat{v}^{l}|u_{n_{1}}\rangle$, $\omega_{nn_{1}}=\varepsilon
_{n}-\varepsilon_{n_{1}}$, and $|u_{n}\rangle$ is the periodic part of the
Bloch wave with band energy $\varepsilon_{n}$. The summation over repeated
Cartesian indices is implied henceforth.

Meanwhile, the local $\boldsymbol{s}$-generation rate due to $\boldsymbol{s}$
nonconservation takes the form of
\begin{equation}
\boldsymbol{\tau}\left(  \boldsymbol{r}\right)  =-\boldsymbol{\nabla}%
\cdot\boldsymbol{J}^{\boldsymbol{\tau}}\left(  \boldsymbol{r}\right)  ,\text{
\ }J^{i\boldsymbol{\tau}}=D^{i\boldsymbol{\tau}}-\partial_{l}%
Q^{il\boldsymbol{\tau}}, \label{torque}%
\end{equation}
where we have used the fact that the value of $\boldsymbol{\tau}$ in the
zeroth order of spatial gradients, $\int f_{n}\boldsymbol{\tau}_{n}$,
vanishes, and
\begin{align}
D^{i\boldsymbol{\tau}}  &  =\int(f_{n}d_{n}^{i\boldsymbol{\tau}}+g_{n}%
\Omega_{n}^{i\boldsymbol{\tau}}),\text{ }\nonumber\\
Q^{il\boldsymbol{\tau}}  &  =\int(f_{n}q_{n}^{il\boldsymbol{\tau}}+g_{n}%
\chi_{n}^{il\boldsymbol{\tau}}).
\end{align}
Here $D^{i\boldsymbol{\tau}}$ is the equilibrium $\boldsymbol{\tau}$-dipole
density, which involves not only the $\boldsymbol{\tau}$-dipole moment
$d_{n}^{i\boldsymbol{\tau}}$\ of each wave packet but also a corresponding
Berry curvature $\Omega_{n}^{i\boldsymbol{\tau}}$. The latter was overlooked
previously, but, as will be shown later, plays an important role in ensuring
the vanishing net flow of the conserved current. $Q^{il\boldsymbol{\tau}}$ is
essentially the $\boldsymbol{\tau}$-quadrupole density
$Q^{(il)\boldsymbol{\tau}}=(Q^{il\boldsymbol{\tau}}+Q^{li\boldsymbol{\tau}%
})/2$ and contributes to the $\boldsymbol{s}$-generation rate as
$-\partial_{i}\partial_{l}Q^{il\boldsymbol{\tau}}$. Here $q_{n}%
^{il\boldsymbol{\tau}}=\operatorname{Re}\langle\frac{1}{2}\left(  \hat
{r}-r_{c}\right)  ^{i}\left(  \hat{r}-r_{c}\right)  ^{l}\hat{\boldsymbol{\tau
}}\rangle$ is the $\boldsymbol{\tau}$-quadrupole moment and $\chi
_{n}^{il\boldsymbol{\tau}}=\partial_{k_{i}}d_{n}^{l\boldsymbol{s}%
}-2\operatorname{Im}\sum_{n_{1}\neq n}v_{nn_{1}}^{l}d_{n_{1}n}%
^{i\boldsymbol{\tau}}/\omega_{nn_{1}}^{2}$ is the $\boldsymbol{\tau}$-dipole
polarizability of a bulk semiclassical electron, with $d_{n_{1}n}%
^{i\boldsymbol{\tau}}$ having the meaning of interband $\boldsymbol{\tau}%
$-dipole moment \cite{interband dipole}. In literatures on the conserved spin
current of Bloch electrons, $J^{i\boldsymbol{\tau}}$ is termed as the torque
($\boldsymbol{\hat{\tau}}$ is the torque operator in this context) dipole spin
current. Our theory shows that this understanding is inaccurate as the torque
quadrupole density is also involved. In fact, it is this quadrupole
contribution that cancels out the non-circulating part of the conventional
$\boldsymbol{s}$-current and makes the current conserved.

According to the above evaluation, $\boldsymbol{J}^{\boldsymbol{\tau}}$ can be
deemed as a $\boldsymbol{s}$-current arising from the nonconservation of
$\boldsymbol{s}$ and we can inspect if
\begin{equation}
\boldsymbol{\mathcal{J}}^{\boldsymbol{s}}=\boldsymbol{J}^{\boldsymbol{s}%
}+\boldsymbol{J}^{\boldsymbol{\tau}} \label{current}%
\end{equation}
is a conserved current density. Generally speaking, a conserved current takes
the form of $\boldsymbol{\mathcal{J}}^{\boldsymbol{s}}=\boldsymbol{\mathcal{J}%
}_{\text{un}}^{\boldsymbol{s}}+\boldsymbol{\nabla}\times\boldsymbol{M}%
^{\boldsymbol{s}}$, where $\boldsymbol{\mathcal{J}}_{\text{un}}%
^{\boldsymbol{s}}$ is the equilibrium current in the uniform case, and
$\boldsymbol{M}^{\boldsymbol{s}}$ is referred to as the $\boldsymbol{s}$
orbital magnetization by analogy to the charge orbital magnetization. The
developed semiclassical theory enables us to show the conserving nature of
$\boldsymbol{\mathcal{J}}^{\boldsymbol{s}}$ directly and determine
$\boldsymbol{\mathcal{J}}_{\text{un}}^{\boldsymbol{s}}$ and $\boldsymbol{M}%
^{\boldsymbol{s}}$.

\emph{{\color{blue} Orbital magnetization and \emph{vanishing net flow}.}}---
Due to the peculiarity of the $\boldsymbol{s}$-generation rate operator, we
get $d_{n}^{i\boldsymbol{\tau}}=-J_{n}^{i\boldsymbol{s}}+v_{n}^{i}%
\boldsymbol{s}_{n}$ and $\Omega_{n}^{i\boldsymbol{\tau}}=\partial_{k_{i}%
}\boldsymbol{s}_{n}$. One finds that the $\boldsymbol{\tau}%
$-dipole moment quantifies the deviation of the conventional $\boldsymbol{s}%
$-current from the classical form of $\boldsymbol{s}$-current due to
$\boldsymbol{s}$ nonconservation, and the $\boldsymbol{\tau}$-dipole density
$D^{i\boldsymbol{\tau}}$ cancels out the conventional $\boldsymbol{s}$-current
$\int f_{n}J_{n}^{i\boldsymbol{s}}$. We thus uncover the first important
property of the current defined in Eq. (\ref{current}) that it vanishes in
uniform equilibrium
\begin{equation}
\boldsymbol{\mathcal{J}}_{\text{un}}^{\boldsymbol{s}}=0. \label{net}%
\end{equation}
This result means that the net flow of $\boldsymbol{\mathcal{J}}%
^{\boldsymbol{s}}$ through any cross section of a sample vanishes, which is a
necessary character of a current that can describe transport.

Next, after tedious manipulations of the $\boldsymbol{\tau}$-quadrupole
density \cite{supp}, we get $\boldsymbol{\mathcal{J}}^{\boldsymbol{s}%
}=\boldsymbol{\nabla}\times\boldsymbol{M}^{\boldsymbol{s}}$ with the
$\boldsymbol{s}$ orbital magnetization given by \cite{gauge}%
\begin{equation}
\boldsymbol{M}^{\boldsymbol{s}}=\int(f_{n}\boldsymbol{m}_{n}^{\boldsymbol{s}%
}+g_{n}\mathbf{\Omega}_{n}\boldsymbol{s}_{n}).\label{SOM}%
\end{equation}
Here $\boldsymbol{m}_{n}^{\boldsymbol{s}}=\frac{1}{2}\operatorname{Re}%
\sum_{n_{1}\neq n}\mathbf{\mathcal{A}}_{nn_{1}}\times\boldsymbol{J}_{n_{1}%
n}^{\boldsymbol{s}}+\frac{1}{2}\boldsymbol{d}_{n}^{\boldsymbol{s}}%
\times\boldsymbol{v}_{n}$ accounts for the $\boldsymbol{s}$ orbital magnetic
moment carried by each wave packet, with $\mathbf{\mathcal{A}}_{nn_{1}}$ being
the $\boldsymbol{k}$-space interband Berry connection, and$\ \mathbf{\Omega
}_{n}$ is the $\boldsymbol{k}$-space\ Berry curvature. If $\boldsymbol{s}$ is
replaced by charge $e$ of Bloch electrons, $\boldsymbol{M}^{\boldsymbol{s}}$
reduces to the charge orbital magnetization \cite{Xiao2006}. To see the
orbital nature of $\boldsymbol{M}^{\boldsymbol{s}}$ for nonconserved
$\boldsymbol{s}$, we inspect the content of $\boldsymbol{m}_{n}%
^{\boldsymbol{s}}$. The first term of $\boldsymbol{m}_{n}^{\boldsymbol{s}}%
$\ is equal to $\langle\frac{1}{2}\left(  \hat{\boldsymbol{r}}-\boldsymbol{r}%
_{c}\right)  \times\boldsymbol{\hat{J}}^{\boldsymbol{s}}\rangle$, which is the
antisymmetric part of the dipole moment of $\boldsymbol{J}^{\boldsymbol{s}}$
and means the circulation of the conventional $\boldsymbol{s}$-current due to
the self-rotational motion of a wave packet around its center position. On the
other hand, the second term of $\boldsymbol{m}_{n}^{\boldsymbol{s}}$ signifies
remarkably that the center-of-mass motion of a wave packet with a nonvanishing
$\boldsymbol{s}$-dipole moment $\boldsymbol{d}_{n}^{\boldsymbol{s}}$ induces a
$\boldsymbol{s}$ orbital magnetic moment. This term is thus reminiscent of the
phenomenon in electromagnetism that a moving charge dipole moment results in a
charge orbital magnetic moment.

\emph{{\color{blue} Orbital magnetization in superconductors.}}---Now we apply
the above results to the context of orbital magnetization in superconductors.
Intraband spin-singlet pairing without spin-orbit coupling is assumed for
illustration. As has been shown recently, the semiclassical theory for
Bogoliubov quasiparticle wave packets can be formulated similarly to that for
electrons \cite{Liang2017,Wang2020}. Such a theory can accommodate
slowly-varying perturbations whose length scales are much larger than the
superconducting coherence length. The nonlocal pair potential is treated in
the continuum limit in our consideration, but may also be dealt with assuming
a more general periodic form \cite{Liang2017}. To confirm the utility of the
semiclassical theory, we show in the Supplemental Material \cite{supp} that it
readily leads to the quantum thermal Hall and spin quantum Hall (quantized
spin conductivity in response to a Zeeman gradient) effects predicted by field
theoretical methods in chiral $d+id$ superconductors
\cite{Fisher1999,Read2000}.

The conventional charge current carried by a mean-field quasiparticle is
represented by the anticommutator of charge and velocity operators, where the
charge operator is $e$ times the third Pauli matrix $\hat{\sigma}^{z}$ in the
electron-hole space \cite{Liang2017,Nambu2009}: $\boldsymbol{\hat{s}}%
=e\hat{\sigma}^{z}$. It is simply given by $e\partial_{\boldsymbol{k}}%
\xi_{\eta\boldsymbol{k}}$, where $\xi_{\eta}$ is the energy of an electron
relative to its chemical potential, with $\eta$ as the Bloch band index
\cite{Lee2006}. Such a conventional current carried by an ensemble of
mean-field quasiparticles, namely $\boldsymbol{J}^{\boldsymbol{s}}$, may not
vanish at equilibrium in uniform systems without time-reversal and inversion symmetries.

The source term $\boldsymbol{\tau}=-\boldsymbol{\nabla}\cdot\boldsymbol{J}%
^{\boldsymbol{\tau}}$ for this conventional current arises from the fact that
the mean-field quasiparticles are not eigenstates of charge and embodies the
coupling between mean-field condensates and quasiparticles
\cite{Nambu1960,BTK1982,Kivelson2012,Wu2017}. Thus $\boldsymbol{J}%
^{\boldsymbol{\tau}}$ serves as a condensate backflow, which makes the current
conserved and cancels out the net flow of the conventional current carried by
mean-field quasiparticles at equilibrium [Eq. (\ref{net})].

The orbital magnetization reads
\begin{equation}
\boldsymbol{M}=\int(f_{n}\boldsymbol{m}_{n}+e\rho_{n}g_{n}\mathbf{\Omega}%
_{n}). \label{OM}%
\end{equation}
Here and hereafter we omit the superscript $\boldsymbol{s}$ in charge current
related quantities. $\rho_{n}=\sigma_{n}^{z}=\left\vert \mu_{n}\right\vert
^{2}-\left\vert \nu_{n}\right\vert ^{2}=\sigma\rho_{\eta}^{0}$ is the charge
($e\rho_{n}$) carried by a mean-field quasiparticle, for which the band index
$n=\left(  \eta,\sigma\right)  $, with $\sigma=\pm$ denoting the Bogoliubov
bands. $[\mu_{n\boldsymbol{k}},\nu_{n\boldsymbol{k}}]^{T}$ is the Bogoliubov
wave function, $\rho_{\eta}^{0}=\xi_{\eta}/|\varepsilon_{\eta\sigma}|$, and
$\varepsilon_{\eta\sigma}=\sigma\sqrt{\xi_{\eta}^{2}+\left\vert \Delta_{\eta
}\right\vert ^{2}}$. The orbital magnetic moment of each quasiparticle is%
\begin{equation}
\boldsymbol{m}_{n}=\frac{1}{2}\operatorname{Re}\sum_{n_{1}\neq n}%
\mathbf{\mathcal{A}}_{nn_{1}}\times\boldsymbol{J}_{n_{1}n}+\frac{1}%
{2}\boldsymbol{d}_{n}\times\boldsymbol{v}_{n},
\end{equation}
where
\begin{equation}
\boldsymbol{d}_{n}=e{{\operatorname{Re}\langle\left(  \boldsymbol{\hat{r}%
}-\boldsymbol{r}_{c}\right) \hat{\sigma}^{z}\rangle}}=\frac{e}{2}%
[(\rho_{\eta}^{0})^{2}-1]\partial_{\boldsymbol{k}}\theta_{\eta}%
\end{equation}
is the charge dipole moment of a quasiparticle wave packet, with $\theta
_{\eta}=\arg\Delta_{\eta}$ being the phase of the superconducting gap function
in $\boldsymbol{k}$-space. Nonzero dipole moment signifies that the charge
distribution on a Bogoliubov wave packet is not centered at the probability
center of the wave packet. This charge dipole is a basic property of
quasiparticles in unconventional pairing superconductors, and has also been
identified recently from a different method \cite{Wang2020}.

Our account of the orbital magnetization is limited to strongly type-II
superconductors where the penetration length is irrelevant (much larger than
the length scale considered), such as in two dimensional (2D) systems. In such
a case the screening current can be neglected and the consideration of
magnetization is simplified. In more complicated cases where the Meissner
screening must be taken into account, a theory incorporating the Coulomb
interaction is needed. This is left for a future work.

The orbital magnetization [Eq. (\ref{OM})] is interpreted as a sum of local
and global charge circuits of quasiparticle wave packets. The first term of
$\boldsymbol{m}_{n}$\ is the circulation of the conventional charge current
about the wave-packet center, whereas the second term signifies an orbital
magnetic moment induced by a travelling charge dipole (this latter mechanism
does not show up for Bloch electrons since their charge center coincides with
the wave-packet center). Therefore, $\boldsymbol{m}_{n}$ can be understood as
the local circuit accompanying a moving wave packet. Meanwhile, the Berry
curvature term in $\boldsymbol{M}$ results from the global circuit due to the
center-of-mass motion of wave packets \cite{Xiao2010,Gradhand2020}.

A vanishing orbital magnetization $\boldsymbol{M}=\boldsymbol{0}$ is predicted
for the case of geometrically trivial electronic bands. In this case one has a
particle-hole symmetric two-band model with Bogoliubov band index $\sigma=\pm
$, for which the Berry curvature $\mathbf{\Omega}_{\sigma}=-\sigma
\partial_{\boldsymbol{k}}\rho^{0}\times\partial_{\boldsymbol{k}}\theta/2$
stems from unconventional superconducting pairing. In addition, the first term
of $\boldsymbol{m}_{n}$\ vanishes due to the particle-hole symmetry. It is
then apparent that the statistical sum of local circuits cancels exactly the
global circuit. On the other hand, such a null result is not anticipated for
the case of geometrically nontrivial Bloch bands.

\emph{{\color{blue} Transport in terms of the conserved current.}}---Next we
turn to the nonequilibrium case with a weak uniform electric field
$\boldsymbol{E}$, chemical potential gradient $\boldsymbol{\nabla}\mu$ and
temperature gradient $\boldsymbol{\nabla}T$. The electric field and the
chemical potential gradient are only considered in the context of Bloch
electrons (in the perspective of transport, an electric field for Bloch
electrons corresponds to a Zeeman gradient applied to a spin-singlet
superconductor without spin-orbit coupling \cite{Fisher1999,Read2000}). The
first observation is that the steady-state linear response of {$\int
f_{n}\boldsymbol{\tau}_{n}$} vanishes if $\boldsymbol{\hat{s}}$ commutes with
the position operator. Showing this conclusion entails only a standard linear
response analysis in disordered systems \cite{supp,Sugimoto2006}. Thus, at
nonequilibrium the current defined by Eq. (\ref{current}) is still conserved.

The intrinsic transport current is found by subtracting the circulating
magnetization one: $\boldsymbol{\mathcal{J}}^{\boldsymbol{s}}%
-\boldsymbol{\nabla}\times\boldsymbol{M}^{\boldsymbol{s}}$, and is quantified
{by the sum of the $\boldsymbol{s}$-Berry curvature and }$\boldsymbol{\tau}%
$-dipole polarizability: $\Omega_{n}^{li\boldsymbol{s}}+\chi_{n}%
^{il\boldsymbol{\tau}}=\partial_{k_{l}}d_{n}^{i\boldsymbol{s}}+\Omega_{n}%
^{li}\boldsymbol{s}_{n}$, which turns out to be expressed by the
$\boldsymbol{s}$-dipole moment and $\boldsymbol{k}$-space Berry curvature
tensor $\Omega_{n}^{il}$. The result is%
\begin{equation}
\mathcal{J}_{\text{in}}^{i\boldsymbol{s}}=\sigma_{\text{in}}^{il\boldsymbol{s}%
}(E_{l}-\partial_{l}\mu/e)-\alpha_{\text{in}}^{il\boldsymbol{s}}\partial_{l}T,
\end{equation}
with%
\begin{align}
\sigma_{\text{in}}^{il\boldsymbol{s}}  &  =e\int f_{n}(\Omega_{n}%
^{li}\boldsymbol{s}_{n}+\partial_{k_{l}}d_{n}^{i\boldsymbol{s}}),\nonumber\\
\alpha_{\text{in}}^{il\boldsymbol{s}}  &  =-\int\partial g_{n}/\partial
T(\Omega_{n}^{li}\boldsymbol{s}_{n}+\partial_{k_{l}}d_{n}^{i\boldsymbol{s}}).
\label{SHC}%
\end{align}
Here $-\partial g_{n}/\partial T$ is the state resolved entropy density, and
the Einstein and Mott relations are ensured. In superconductors, as has been
mentioned, the charge current driven by a temperature gradient is considered
for strongly type-II case. In the case of conserved spin current of Bloch
electrons, the above result for $\sigma_{\text{in}}^{il\boldsymbol{s}}$
unveils the band geometric origin of the intrinsic spin conductivity obtained
in the quantum theory \cite{Shi2006}, and $\alpha_{\text{in}}%
^{il\boldsymbol{s}}$/$\sigma_{\text{in}}^{il\boldsymbol{s}}$ is consistent
with the intrinsic transport thermal/charge current driven by a Zeeman
gradient \cite{Xiao2020EM}.

\emph{{\color{blue} Spin transport of Bloch electrons.}}---In this context
$\boldsymbol{\hat{s}}$ is spin, and Eqs. (\ref{current}) -- (\ref{SOM}) and
(\ref{SHC}) solve all the problems mentioned in the Introduction. Moreover, we
find that even if the considered spin component is not conserved, the spin
conductivity in insulators is of purely Hall type $\boldsymbol{\mathcal{J}%
}^{\boldsymbol{s}}=\boldsymbol{E}\times\boldsymbol{\sigma}_{H}^{\boldsymbol{s}%
}$, a characteristic that is not shared by the conventional spin current. The
Hall conductivity is related to the spin orbital magnetization:
\begin{equation}
\boldsymbol{\sigma}_{H}^{\boldsymbol{s}}=-e\int\boldsymbol{\Omega}%
_{n}\boldsymbol{s}_{n}=e\frac{\partial\boldsymbol{M}^{\boldsymbol{s}}%
}{\partial\mu}.
\end{equation}

Our theory also sheds new light on disorder influenced spin transport in terms
of $\boldsymbol{\mathcal{J}}^{\boldsymbol{s}}$ in metals. Currents stemming
from the $\boldsymbol{E}$-field driven off-equilibrium occupation function
read$\ J_{\text{c}}^{i\boldsymbol{s}}=\int\delta f_{n}J_{n}^{i\boldsymbol{s}}$
and $J_{\text{c}}^{i\boldsymbol{\tau}}=\int\delta f_{n}d_{n}%
^{i\boldsymbol{\tau}}$ \cite{Murakami2015}. The resulting conserved
current\ is unexpectedly simple as if the spin were conserved:
\begin{equation}
\mathcal{J}_{\text{c}}^{i\boldsymbol{s}}=\int\delta f_{n}v_{n}^{i}%
\boldsymbol{s}_{n}. \label{extrinsic}%
\end{equation}
It may have transverse components in anisotropic systems even if $\delta
f_{n}$ is determined by the conventional Boltzmann equation with the first
Born scattering rate \cite{Karel2009}. $\mathcal{J}_{\text{c}}%
^{i\boldsymbol{s}}$ is time-reversal odd and is relevant to the study of
magnetic spin Hall effect \cite{Yan2017,Mertig2020}.

The spin current arises from intra-scattering semiclassics as well. This is a
conceptual generalization of the side jump physics rooted in the coordinate
shift of an electron wave packet during scattering
\cite{Sinitsyn2006,Sinitsyn2008}. When the transported quantity is
nonconserved, the key ingredient is replaced by the shift of a $\boldsymbol{s}%
$-coordinate ($\boldsymbol{\hat{r}}^{\boldsymbol{s}}=\boldsymbol{\hat{s}%
\hat{r}}$) upon a scattering $n\boldsymbol{k}\rightarrow n^{\prime
}\boldsymbol{k}^{\prime}$. We find \cite{supp} $\delta\boldsymbol{r}%
^{\boldsymbol{s}}=\delta_{\text{dp}}\boldsymbol{r}^{\boldsymbol{s}}%
+\delta_{\text{sj}}\boldsymbol{r}^{\boldsymbol{s}}$, where $(\delta
_{\text{sj}}\boldsymbol{r}^{\boldsymbol{s}})_{\boldsymbol{k}^{\prime
}\boldsymbol{k}}^{n^{\prime}n}=\boldsymbol{s}_{n^{\prime}\boldsymbol{k}%
^{\prime}}\boldsymbol{\mathcal{A}}_{n^{\prime}\boldsymbol{k}^{\prime}%
}-\boldsymbol{s}_{n\boldsymbol{k}}\boldsymbol{\mathcal{A}}_{n\boldsymbol{k}%
}-(\boldsymbol{s}_{n^{\prime}\boldsymbol{k}^{\prime}}\partial_{\boldsymbol{k}%
^{\prime}}+\boldsymbol{s}_{n\boldsymbol{k}}\partial_{\boldsymbol{k}})\arg
V_{\boldsymbol{k}^{\prime}\boldsymbol{k}}^{n^{\prime}n}$ is reminiscent of the
coordinate shift \cite{Sinitsyn2006} and reduces to the latter for conserved
$\boldsymbol{s}$, with $\arg V$ being the phase of the scattering matrix
element, whereas
\begin{equation}
(\delta_{\text{dp}}\boldsymbol{r}^{\boldsymbol{s}})_{\boldsymbol{k}^{\prime
}\boldsymbol{k}}^{n^{\prime}n}=\boldsymbol{d}_{n^{\prime}\boldsymbol{k}%
^{\prime}}^{\boldsymbol{s}}-\boldsymbol{d}_{n\boldsymbol{k}}^{\boldsymbol{s}%
}\label{sj-1}%
\end{equation}
is the intra-scattering change of the $\boldsymbol{s}$-dipole moment. Both
parts of $\delta\boldsymbol{r}^{\boldsymbol{s}}$ are independent of the phase
choice of Bloch functions, implying two semiclassical contributions to the
spin current: $\boldsymbol{\mathcal{J}}_{\text{dp/sj}}^{\boldsymbol{s}}%
=\int[d\boldsymbol{k}]\delta f_{n\boldsymbol{k}}\boldsymbol{\mathcal{J}%
}_{\text{dp/sj},n\boldsymbol{k}}^{\boldsymbol{s}}$ \cite{note}, where
$\boldsymbol{\mathcal{J}}_{\text{dp/sj},n\boldsymbol{k}}^{\boldsymbol{s}}%
=\int[d\boldsymbol{k}^{\prime}]P_{\boldsymbol{k}^{\prime}\boldsymbol{k}%
}^{n^{\prime}n}(\delta_{\text{dp/sj}}\boldsymbol{r}^{\boldsymbol{s}%
})_{\boldsymbol{k}^{\prime}\boldsymbol{k}}^{n^{\prime}n}$ is the current
carried by each electron and quantifies the accumulation of $\delta
\boldsymbol{r}^{\boldsymbol{s}}$ upon possible scattering events per unit
time, with $P_{\boldsymbol{k}^{\prime}\boldsymbol{k}}^{n^{\prime}n}$ being the
first Born scattering rate \cite{Sinitsyn2008}.

Appearing only for nonconserved $\boldsymbol{s}$, $\delta_{\text{dp}%
}\boldsymbol{r}^{\boldsymbol{s}}$ is noticeable as it originates from
scattering but its expression is independent of scattering. It is the
difference of a \textquotedblleft function of state\textquotedblright%
\ $\boldsymbol{d}_{n\boldsymbol{k}}^{\boldsymbol{s}}$\ upon a scattering
process. This feature has a remarkable influence. By utilizing the relevant
Boltzmann equation in the lowest Born order \cite{Sinitsyn2008}, we find that
$\boldsymbol{\mathcal{J}}_{\text{dp}}^{\boldsymbol{s}}=-eE_{l}\int
f_{n}\partial_{k_{l}}\boldsymbol{d}_{n}^{\boldsymbol{s}}$, cancelling out the
$\boldsymbol{s}$-dipole term of the intrinsic $\boldsymbol{s}$-current [Eq.
(\ref{SHC})] regardless of the specific form of a weak disorder potential.
Hence the band structure dictated current is of Hall type in metals and is
connected to the Berry curvature:%
\begin{equation}
\boldsymbol{\mathcal{J}}_{\text{B}}^{\boldsymbol{s}}=\boldsymbol{\mathcal{J}%
}_{\text{in}}^{\boldsymbol{s}}+\boldsymbol{\mathcal{J}}_{\text{dp}%
}^{\boldsymbol{s}}=-e\boldsymbol{E}\times\int f_{n}\boldsymbol{\Omega}%
_{n}\boldsymbol{s}_{n}.
\end{equation}

Gathering the results, the transport conserved current in the semiclassical
theory is $\boldsymbol{\mathcal{J}}^{\boldsymbol{s}}=\boldsymbol{\mathcal{J}%
}_{\text{B}}^{\boldsymbol{s}}+\boldsymbol{\mathcal{J}}_{\text{sj}%
}^{\boldsymbol{s}}+\boldsymbol{\mathcal{J}}_{\text{c}}^{\boldsymbol{s}}$. As
an application, we reveal that the puzzling null results of the conserved spin
Hall conductivity in 2D Rashba-type models within the weak disorder regime
\cite{Sugimoto2006,Mandal2008,note-weakdisorder} is due to $s_{n}^{z}=0$ in
such models. The $C_{\infty v}$ symmetry forbids spin conductivities with spin
polarization along in-plane directions, and $\boldsymbol{\mathcal{J}%
}_{\text{B/sj/c}}^{\boldsymbol{s}}=0$ for $s=s^{z}$. In particular, the
intrinsic contribution $\boldsymbol{\mathcal{J}}_{\text{in}}^{\boldsymbol{s}}%
$, which reproduces the results of the linear response theory in clean systems
\cite{Shi2006}, is canceled out by $\boldsymbol{\mathcal{J}}_{\text{dp}%
}^{\boldsymbol{s}}$.

\emph{{\color{blue} Concluding remarks.}}---The theoretical framework
developed may also be useful in other subjects of interest, such as the layer
pseudospin Hall effect in twisted bilayers \cite{Yu2019} and spin Nernst effect
in spin-triplet or spin-orbit coupled superconductors with Bogoliubov Fermi
surfaces \cite{Timm2017}. Besides, to generalize the theory to magnon and
phonon systems described by bosonic Bogoliubov-de Gennes Hamiltonians is of
importance for studying the spin Nernst effect of magnons in a noncollinear
antiferromagnetic insulators \cite{Mook2019,Li2020} and that due to
magnon-phonon interactions in collinear ferrimagnets \cite{Park-1} as well as
the phonon angular momentum Hall effect \cite{Park-2}. In all these subjects
there is the issue of the conserved current of nonconserved quantities. In
addition, in superconductors the present steady-state semiclassical theory
need be extended to include time derivatives and gauge fields in order to
describe gauge-invariant coupled dynamics of quasiparticles and condensate in
the presence of electromagnetic fields \cite{Kita2001,Wu2017}.

\begin{acknowledgments}
We thank Yang Zhang, Zhi Wang, Yinhan Zhang, Liang Dong, Wang Yao, Yang Gao and Tianlei Chai for stimulating discussions.
This work is supported by NSF (EFMA-1641101) and Welch Foundation (F-1255).
\end{acknowledgments}


\begin{thebibliography}{99}
\bibitem {interband dipole}The expression of $d_{n_{1}n}^{i\boldsymbol{\tau}}$
($n_{1}\neq n$) is provided here (the derivation is presented in Ref.
\cite{supp}) for interested readers:
\begin{align*}
d_{n_{1}n}^{i\boldsymbol{\tau}}  &  =\frac{1}{2}(i\partial_{k_{i}}%
+\mathcal{A}_{n_{1}}^{i}-\mathcal{A}_{n}^{i})\boldsymbol{\tau}_{n_{1}n}%
-\sum_{n_{2}\neq n}\delta_{n_{2}n_{1}}\frac{i\boldsymbol{\tau}_{n_{2}n}%
v_{n}^{i}}{\varepsilon_{n}-\varepsilon_{n_{2}}}\\
&  +\frac{1}{2}\sum_{n_{2}\neq n_{1}}\mathcal{A}_{n_{1}n_{2}}^{i}%
\boldsymbol{\tau}_{n_{2}n}+\frac{1}{2}\sum_{n_{2}\neq n}\boldsymbol{\tau
}_{n_{1}n_{2}}\mathcal{A}_{n_{2}n}^{i},
\end{align*}
where $\mathcal{A}_{n}^{i}$ is the $\boldsymbol{k}$-space Berry connection. It
can be extended to involve the $\boldsymbol{\tau}$-dipole moment
$d_{n}^{i\boldsymbol{\tau}}$ when $n_{1}=n$, hence can be deemed as the
interband element of the $\boldsymbol{\tau}$-dipole moment.
\end{thebibliography}
\end{document}